\documentclass[aps,reprint,twocolumns,pre,superscriptaddress,floatfix,notitlepage,nofootinbib]{revtex4-1}
\usepackage{amssymb,amsmath,amsfonts} 
\usepackage{mathtools}
\usepackage{physics}
\usepackage{array}
\usepackage{graphicx,epsfig,xcolor}
\usepackage[colorlinks=true,citecolor=blue,linkcolor=red]{hyperref}
\usepackage{newtxtext,newtxmath}

\begin{document}

\title{Chaos in a deformed Dicke model}

\author{\'{A}ngel L. Corps}
\email[]{corps.angel.l@gmail.com}
\affiliation{Instituto de Estructura de la Materia, IEM-CSIC, Serrano 123, E-28006 Madrid, Spain}
\affiliation{Departamento de Estructura de la Materia, F\'{i}sica T\'{e}rmica y Electr\'{o}nica \& Grupo Interdisciplinar de Sistemas Complejos (GISC), Universidad Complutense de Madrid, Av. Complutense s/n, E-28040 Madrid, Spain}

\author{Rafael A. Molina}
\email[]{rafael.molina@csic.es}
\affiliation{Instituto de Estructura de la Materia, IEM-CSIC, Serrano 123, E-28006 Madrid, Spain}

\author{Armando Rela\~{n}o}
\email[]{armando.relano@fis.ucm.es}
\affiliation{Departamento de Estructura de la Materia, F\'{i}sica T\'{e}rmica y Electr\'{o}nica \& Grupo Interdisciplinar de Sistemas Complejos (GISC), Universidad Complutense de Madrid, Av. Complutense s/n, E-28040 Madrid, Spain}

\date{\today}

\begin{abstract}
The critical behavior in an important class of excited state quantum phase transitions is signaled by the presence of a new constant of motion only at one side of the critical energy. We study the impact of this phenomenon in the development of chaos in   
a modified version of the paradigmatic Dicke model of quantum optics, in which a perturbation is added that breaks the parity symmetry. Two asymmetric energy wells appear in the semiclassical limit of the model, whose consequences are studied both in the classical and in the quantum cases. Classically, Poincar\'{e} sections reveal that the degree of chaos not only depends on the energy of the initial condition chosen, but also on the particular energy well structure of the model. In the quantum case, Peres lattices of physical observables show that the appearance of chaos critically depends on 
the quantum conserved number provided by this constant of motion. The conservation law defined by this constant is shown to allow for the coexistence between chaos and regularity at the same energy. 
We further analyze the onset of chaos in relation with an additional conserved quantity that the model can exhibit. 
\end{abstract}

\maketitle

\section{Introduction}
Symmetries have famously played an important role throughout physics, and this is no different within the field of quantum chaos. Quantum chaos as a whole has been shaped to an exceptional extent by the work of Fritz Haake, whose monograph \cite{Haakebook} continues to be one of the main references on the subject. 

The origins of quantum chaos can be arguably traced back to the 1950s, when Wigner introduced the random matrix theory (RMT) in an attempt to understand the lack of states observed in certain nuclear resonances \cite{Wigner1955}. At that time the RMT was only being used as a mathematical tool to describe a complicated system such as the atomic nucleus. But even still in the early days of quantum chaos it became clear that two ingredients were needed to understand the features of quantum systems whose classical analogue exhibits chaotic dynamics: a combination of the RMT \cite{Mehta1990} and the symmetries satisfied by the system under consideration. In 1984, in their seminal work, Bohigas, Giannoni and Schmit (BGS) \cite{Bohigas1984} put forward the conjecture that the level fluctuations of quantally chaotic systems are universal and given by the RMT. Depending on the particular symmetries of the system, level statistics would belong to the universality class of one of the three classical random ensembles: the Gaussian, unitary and symplectic ensembles \cite{Guhr1998}. This is in stark contrast to generic quantum integrable systems, whose eigenlevels are statistically uncorrelated and belong to the class of the Poisson point process, as put forward by Berry and Tabor already in 1977 \cite{Berry1977}. Examples are known in the literature where this correspondence does not seem to be fulfilled \cite{Mailoud2021,Wu1990,Benet2003,Benet2003b}.  Quantum chaos as we understand it today is deeply rooted in a connection between classical periodic orbit theory, developed by Gutzwiller \cite{Gutzwiller1971}, and the universal statistics of the eigenlevels of certain random matrices. An understanding of this elegant and simple idea, based on the notion of universality of level fluctuations, is a big part of the scientific legacy left by Haake. Expanding on earlier works by Hannay, Ozorio de Almeida, Sieber and Richter \cite{Hannay1984,Sieber2001,Sieber2002}, Haake and his collaborators provided a semiclassical proof of the BGS conjecture in 2004 \cite{Muller2004}. They showed that universality in quantum chaos follows from the observation that in the semiclassical limit the specific features of a system play no role. This was later followed by a series of papers where they laid the foundations of a full periodic-orbit theory for the correlations in the level statistics of quantum chaotic systems \cite{Muller2005,Heusler2007,Muller2009}. Quantum chaos has evolved today to include systems for which there exists no semiclassical limit \cite{Montambaux1993}. In this case, a theory for what quantum chaos may mean beyond compliance with RMT universal results is still lacking. Additionally, basic, fundamental phenomena such as quantum thermalization 
are directly linked to quantum chaotic eigenstates \cite{Santos2010,Alessio2016}. 

One of the most powerful approaches to quantum chaos, at least when one has direct access to the spectrum, is based precisely on the statistical analysis of the eigenlevels. But even this method is not unsensitive to the number of conserved quantities in a given model. One first needs to desymmetrize the spectrum, i.e., separate the eigenvalues in symmetry subspaces. This is because the eigenlevels belonging to different symmetry sectors are uncorrelated, and thus in mixing them all one can easily miss \cite{Gurevich1956,Bunimovich1979} the paradigmatic feature of quantum chaotic spectra: level correlations. The result would appear to be a significantly less chaotic system. Moreover, depending on the statistic considered, one may still need to go through the so-called unfolding procedure, which is non-trivial and can induce spurious correlations where there are really none \cite{Gomez2002,Corps2021pre}; it can even be impracticable when the sequence of eigenlevels is too short. As a consequence, other ways to detect quantum chaos have been developed, including the structure of the quantum eigenstates, expectation value of generic physical observables, eigenstate localization \cite{Santos2012,Zyczkowski1990,Borgonovi2016,Gomez2011}, to mention a few. 
Here, for reasons that will be explained in the following, we focus on Peres lattices \cite{Peres1984}, which characterize pictorically how the quantum numbers associated with the integrability of the model are scrambled as chaoticity increases. 

In this work we focus on a modified version of the Dicke model \cite{Dicke1954}, a prototypical model of quantum optics describing the interaction between $N$ identical two-level atoms (matter) and a single bosonic mode (light). The Dicke model, formally the simplest spin-boson system with more than one atom, is 
a prototypical example of a system exhibiting a normal-superradiant phase transition \cite{Hepp1973,Hepp1973b}, and it also allows for both classical and quantum chaos \cite{Emary2003,Emary2003prl,Kus1985} for suitable values of its parameters. A semiclassical limit for the model can also be obtained \cite{Aguiar1992}, so the quantum-classical correspondence can be readily analyzed. Superradiance (and also superfluorescence) phase transitions of this type greatly benefited from the theory by Haake with Glauber and others \cite{Bonifacio1971,Bonifacio1971b,Haake1972,Haake1979,Haake1978}. Haake's work on the Dicke model includes the famous paper \cite{Altland2012}, where he expanded on the original ideas by Emary and Brandes \cite{Emary2003}, and the peculiar equilibration properties and photon quantum fluctuations of the system were thoroughly analyzed using a Fokker-Planck formalism. Around the same time, a new kind of non-analiticity in the spectrum of certain nuclear collective models had been observed \cite{Cejnar2006,Heinze2006}. This new behavior, acknowledged as a new form of quantum phase transition, termed excited-state quantum phase transition (ESQPT) because it occurred in the high-lying excited states of these models, was considered as an independent phenomenon by Caprio, Cejnar and Iachello for the first time in \cite{Caprio2008}. This phenomenology had been identified mainly in integrable models. It was soon revealed that an ESQPT is not unique to integrable systems as it can also take place in non-integrable models. This was shown in \cite{Perez2011}, where the connections of ESQPTs with the onset of quantum chaos were explored for the first time in the Dicke model. So far, most ESQPTs have been identified in systems with an accessible semiclassical limit \cite{Cejnar2021}. This is because ESQPTs are strong quantum manifestations of the features of the semiclassical phase space and can be obtained by nullifying the classical Hamiltonian flow. 

Many of the consequences of ESQPTs have received great exploration in the past years \cite{Cejnar2021}. How ESQPTs affect the onset of quantum chaos is a question still not fully understood.
In \cite{Corps2021}
a (classical and quantum) constant of motion identifying a large class of ESQPTs was reported. This constant of motion establishes a strong conservation law which is obeyed by the dynamics of systems exhibiting ESQPTs. Focusing on a modified version of the Dicke model where the usual parity symmetry is broken, we explore how this new symmetry affects the development of chaos. 
By contrast with the usual Dicke model \cite{Bastarrachea2014b}, where chaos gradually develops as one goes up in energy, we show that when there exists an asymmetry in the classical phase space with associated ESQPTs, chaos is influenced not only by the energy region of the spectrum considered but also by the new conserved quantum numbers imposed by this constant of motion. 

This paper is organized as follows. Sec. \ref{sec:model} is devoted to the model and its semiclassical limit. In Sec. \ref{subsec:deformed} we review the standard Dicke model of quantum optics and introduce a variation of the model based on a direct coupling to an external bosonic reservoir through a deformation strength. The semiclassical limit is analyzed in Sec. \ref{subsec:semiclassical}. The emergence of chaos in the semiclassical limit is exemplified by means of Poincar\'{e} sections in Sec. \ref{sec:classical}. The development of quantum chaos in the corresponding quantum version of the model is studied in Sec. \ref{sec:quantum}. In particular, in Sec. \ref{subsec:pereslattices} we consider two relevant photonic and atomic observables which reveal a chaotic pattern that depends on the energy well structure of the semiclassical limit. In Sec. \ref{subsec:integrabilitybreaking} we analyze the low-energy region of the spectrum where the system behaves very approximately as an integrable model whose eigenvalues are organized in bands as imposed by an approximate conserved quantity. Finally, we gather the conclusions in Sec. \ref{sec:conclusions}.

\section{Model}\label{sec:model}

\subsection{A deformed Dicke Hamiltonian}\label{subsec:deformed}

One of the most important models to describe light-matter interaction is the Dicke model, introduced nearly 70 years ago \cite{Dicke1954}. The Dicke model describes the coupling between a set of $N$ two-level atoms (spin-1/2 particles) interacting with a single mode electromagnetic field. The strength of the coupling depends on a control parameter, $\lambda$. The model has found an enormous amount of applications in the study of quantum chaos and thermalization \cite{LermaHernandez2019,Bastarrachea2014b,Lobez2016,Perez2011,Kloc2018,Relano2016,Bastarrachea2017,Lobez2021} ergodicity and scarring \cite{Pilatowsky2021,Pilatowsky2021b}, and quantum phase transitions \cite{Bastarrachea2014,Emary2003,Emary2003prl,Puebla2013,Perez2017,Baumann2011}, to cite a few.
In this work, we consider a deformation of the Dicke Hamiltonian where we include a direct coupling to an external bosonic reservoir, 

\begin{equation}\label{eq:hamiltonian}
\hat{\mathcal{H}}=\omega\hat{a}^{\dagger}\hat{a}+\omega_{0}\hat{J}_{z}+\frac{2\lambda}{\sqrt{N}}\hat{J}_{x}(\hat{a}^{\dagger}+\hat{a})+\sqrt{\frac{N}{\omega_{0}j}}\alpha(\hat{a}^{\dagger}+\hat{a}),
\end{equation}
where we set $\hbar\equiv 1$. The parameter $\omega$ is the frequency of the bosonic field, while $\omega_{0}$ represents the constant splitting of the atom eigenlevels. For simplicity, we set $\omega=\omega_{0}=1$ throughout without loss of generality. Here, $\hat{a}^{\dagger}$ and $\hat{a}$ are the usual bosonic creation and annihilation operators, respectively, describing the photonic part of the system. The set $\hat{\mathbf{J}}=(\hat{J}_{x},\hat{J}_{y},\hat{J}_{x})$ are collective pseudo-spin operators corresponding to the $N$ two-level atoms. The total spin operator $\hat{\mathbf{J}}^{2}$ is a conserved quantity, $[\hat{\mathcal{H}},\hat{\mathbf{J}}^{2}]=0$, and its eigenvalues, denoted $j(j+1)$, can be used to separate the Hamiltonian matrix in symmetry sectors. We will work with the maximally symmetric sector defined by its maximum value, $j=N/2$, which includes the ground state, relevant for experimental realizations \cite{Baumann2011}. Finally, $\alpha\in\mathbb{R}$ is the deformation strength. When $\alpha=0$, one trivially recovers the standard Dicke Hamiltonian. As soon as $\alpha\neq0$, the behavior of the system undergoes not only quantitative but also qualitative changes. For example, if $\alpha=0$ the Hamiltonian Eq. \eqref{eq:hamiltonian} admits a discrete $\mathbb{Z}_{2}$ symmetry, called \textit{parity} and given by $\hat{\Pi}\equiv \textrm{exp}\,[i\pi(j+\hat{J}_{z}+\hat{a}^{\dagger}\hat{a})]$. This symmetry allows to separate the eigenstates $\{\left|E_{n,\pm}\right>\}_{n}$ according to its two eigenvalues, $\hat{\Pi}\left|E_{n,\pm}\right>=\pm\left|E_{n,\pm}\right>.$ However, if $\alpha\neq0$, this operator no longer commutes with the Hamiltonian, $[\hat{\mathcal{H}},\hat{\Pi}]\neq0$.
When $\alpha=0$, for $\lambda>\lambda_{c}(\alpha=0)=\sqrt{\omega\omega_{0}}/2$ the Dicke model displays an ESQPT exactly at the reduced energy $\epsilon_{c}=-1$. If $\alpha\neq0$, the situation is more complicated mathematically \cite{Corps2021}.

The Dicke model is non-integrable for all values of $\alpha$ as it does not possess as many conserved quantities as classical degrees of freedom; however, some additional constants of motion can be identified in a certain region of the spectrum, where it does behave very approximately as an integrable model (see below).

\subsection{Semiclassical limit and phase space}\label{subsec:semiclassical}
The Hamiltonian Eq. \eqref{eq:hamiltonian} has an associated classical analogue which describes it in the thermodynamic limit, $N\to\infty$. This mean-field solution can be calculated in several ways. One of them \cite{Pilatowsky2021b} is to take the expectation value of the quantum model $\hat{\mathcal{H}}$ in the tensor product of Glauber-Bloch coherent states $\ket{\textrm{GB}}\equiv \ket{q,p}\otimes\ket{Q,P}$. Here, $\ket{q,p}$ represents Glauber coherent states for the bosonic part, 

\begin{equation}\ket{q,p}=\textrm{exp}\left\{-\frac{j}{4}(q^{2}+p^{2})\right\}\textrm{exp}\left\{\sqrt{\frac{j}{2}}(q+ip)\hat{a}^{\dagger}\right\}\ket{0},\end{equation}
where $\ket{0}$ is the radiation vacuum, while $\ket{Q,P}$ represents Bloch coherent states which apply to the atomic part, 

\begin{equation}\ket{Q,P}=\left(1-\frac{Q^{2}+P^{2}}{4}\right)^{j}\textrm{exp}\left\{\frac{Q+iP}{\sqrt{4-P^{2}-Q^{2}}}\hat{J}_{+}\right\}\ket{j,-j},\end{equation}
being $\ket{j,-j}$ the state with all atoms in the ground-state. Then, the semiclassical model admits the representation

\begin{widetext}\begin{equation}\label{eq:semiclassical}
 H\equiv \frac{\bra{\textrm{GB}}\hat{\mathcal{H}}\ket{\textrm{GB}}}{\omega_{0}j}=\frac{\omega}{2\omega_{0}}(q^{2}+p^{2})+\frac{1}{2}(Q^{2}+P^{2})+\frac{2\lambda q Q}{\omega_{0}}\sqrt{1-\frac{1}{4}(Q^{2}+P^{2})}-1+\sqrt{\frac{2}{\omega_{0}}}\alpha q.
\end{equation}\end{widetext}

\begin{center}
\begin{figure*}[t]
\begin{tabular}{cc}
\hspace*{-0.5cm}\includegraphics[width=0.45\textwidth]{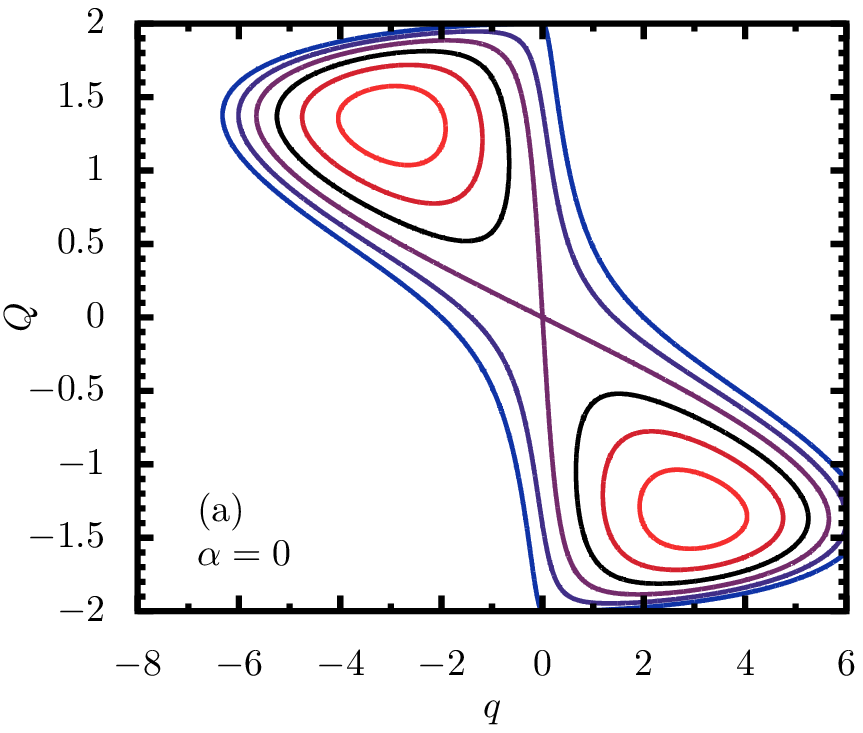} & \includegraphics[width=0.45\textwidth]{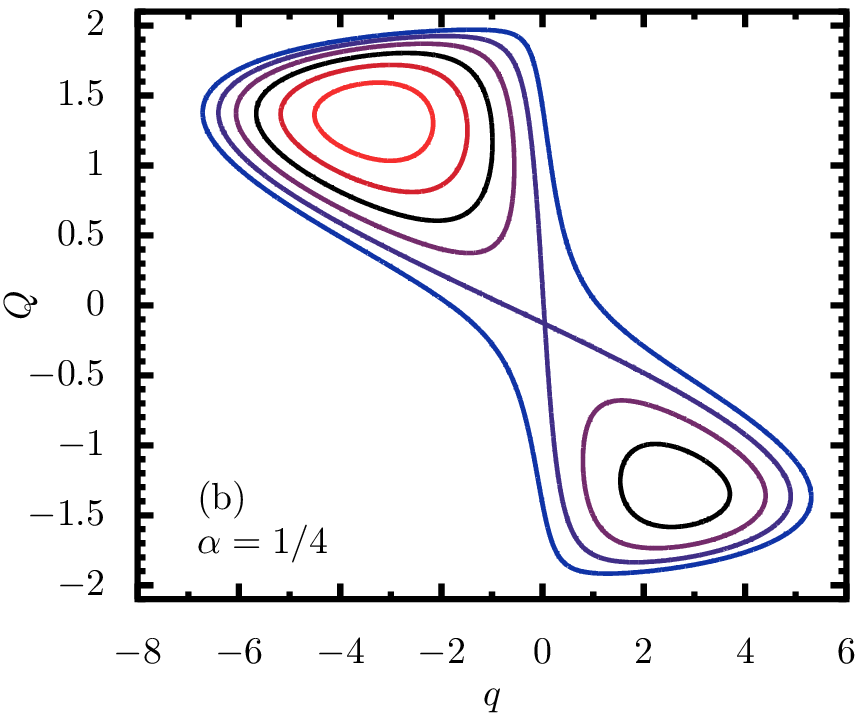}
\end{tabular}
\caption{Projection on the $(q,Q)$ plane of constant energy surfaces for the Hamiltonian function Eq. \eqref{eq:semiclassical} with parameters $\omega_{0}=\omega=1$ and $\lambda=3/2$ for different values of $\alpha$ and reduced energy:  (a) $\alpha=0$,  $\epsilon= -4, -3, -2, -1, 0, 1$; (b) $\alpha=1/4$ and $\epsilon=-4.99,-3.99,-2.99,-1.99,-0.99, 0.01$ (from red to blue).
}
\label{fig:contoursDM}
\end{figure*}
\end{center}

The canonical variables of the classical system are therefore $\mathbf{x}\equiv (q,p;Q,P)\in\mathbb{R}^{4}$, so the semiclassical phase space is four-dimensional, $\mathcal{M}\subseteq\mathbb{R}^{4}$. For convenience, both in the classical and quantum versions of the model we will use the energy scale $\epsilon\equiv E/(\omega_{0}j)$, which enables a direct comparison of the quantum model with the semiclassical limit Eq. \eqref{eq:semiclassical}. The ground-state as well as the various QPTs and ESQPTs can be obtained from the above analytic function. Its phase space coordinates are the critical points $\mathbf{x}^{*}$ such that ${\nabla H}_{\mathbf{x}=\mathbf{x}^{*}}=\mathbf{0}$, and its corresponding energies are readily obtained as $\epsilon^{*}=H(\mathbf{x}^{*})$. When $\alpha=0$ and $\lambda>\lambda_{c}(\alpha=0)$, there exists an analytical expression for these extrema; specifically, they are of the form $\mathbf{x}^{*}=(q^{*},0,Q^{*},0)$ where 

\begin{widetext}
\begin{equation}\label{critsalpha0}
 (q^{*},Q^{*})=(0,0),\left(-\sqrt{\frac{4\lambda^{2}}{\omega^{2}}-\frac{\omega_{0}^{2}}{4\lambda^{2}}},\sqrt{2-\frac{\omega\omega_{0}}{2\lambda^{2}}}\right),\left(\sqrt{\frac{4\lambda^{2}}{\omega^{2}}-\frac{\omega_{0}^{2}}{4\lambda^{2}}},-\sqrt{2-\frac{\omega\omega_{0}}{2\lambda^{2}}}\right).
\end{equation}\end{widetext}
One important aspect of the case $\alpha=0$ is that the classical Hamiltonian is symmetric under $q\to-q$ and $Q\to-Q$; that is, if $(q,p,Q,P)$ is a critical point, so will also be $(-q,p,-Q,P)$ and both correspond to the same energy. For this reason, the second and third critical points in Eq. \eqref{critsalpha0} yield the exact same energy, the ground-state energy. The first critical point is associated with an ESQPT at $\epsilon_{c}=-1$. This scenario is represented in Fig. \ref{fig:contoursDM}(a) where we show the projection of several energy contour surfaces, $\epsilon=H(q,p,Q,P)$, on the $(q,Q)$ plane. Each color is linked to a given energy. We observe that the Hamiltonian does indeed admit two symmetric global energy minima, i.e., the ground-state is degenerate. For $\epsilon\leq\epsilon_{c}$ these two wells remain spatially separated in the phase space. Exactly at the ESQPT criticality, $\epsilon=\epsilon_{c}$, they merge into a single one as can be observed for $\epsilon>\epsilon_{c}$. 

When $\alpha\neq0$, the Hamiltonian function no longer admits the symmetry $H(q,p,Q,P)\to H(-q,p,-Q,P)$, so the above mentioned structure is distorted. In ths case it is not possible to obtain a simple, closed expression for the various critical points and control parameters. In the rest of this work, we will set $\alpha=1/4$ for definiteness. Together with the choice $\omega_{0}=\omega=1$ (in resonance), the ESQPTs appear for couplings $\lambda>\lambda_{c}(\alpha=1/4)=\frac{1}{2}\sqrt{13/16+5\sqrt{17}/16}\approx 0.7247$. We focus on the case $\lambda=3/2>\lambda_{c}$, where the critical points are $\mathbf{x}_{GS}^{*}=(-3.339,0,1.322,0)$, $\mathbf{x}_{1}^{*}=(2.623,0,-1.322,0)$, and $\mathbf{x}_{2}^{*}=(0.045,0,-0.133,0)$.
The corresponding energies are the ground-state energy $\epsilon_{GS}=H(\mathbf{x}_{GS}^{*})=-5.673$ and the critical energies $\epsilon_{c1}=H(\mathbf{x}_{1}^{*})=-3.563$ and $\epsilon_{c2}=H(\mathbf{x}_{2}^{*})=-0.992$.
We can obtain a physical intuition for these extrema in the contours plots in Fig. \ref{fig:contoursDM}(b) for the case $\alpha=1/4$. A consequence of the non-vanishing deformation strength $\alpha$ is that the ground-state is not degenerate anymore; what we obtain are two non-degenerate, non-equivalent minima at different energies. The second minima corresponds to $\mathbf{x}_{1}^{*}$, and it is located at a higher energy than the ground state, $\epsilon_{c1}>\epsilon_{GS}$. Similarly to the case of $\alpha=0$, the energy projection associated to the critical point $\mathbf{x}_{2}^{*}$ at energy $\epsilon_{c2}$ merges both independent energy wells in the $(q,Q)$ plane. 
In the rest of this work we will be interested in this latter case, $\alpha\neq0$.

\section{Chaos in the semiclassical model}\label{sec:classical}

There are several ways to identify the onset of the chaotic regime in the classical realm. A particularly convenient possibility is given by the so-called Poincar\'{e} sections \cite{Reichl1992}, which allow for a very detailed understanding of the underlying classical motion of a Hamiltonian. Obtaining the Poincar\'{e} sections is in principle a rather straightforward procedure, but depending on the equations of motion it can also be computationally expensive. One starts by picking an initial condition $\mathbf{x}(t=0)=(q(0),p(0),Q(0),P(0))$ with a given energy of interest, $\epsilon=H(\mathbf{x}(t=0))$. The classical time-evolution of the initial condition follows the Hamilton equations, which in this case read as follows:  

\begin{equation}
 \frac{\textrm{d}q}{\textrm{d}t}=\frac{\partial H}{\partial p}=\frac{\omega}{\omega_{0}}p,
\end{equation}

\begin{equation}
 \frac{\textrm{d}p}{\textrm{d}t}=-\frac{\partial H}{\partial q}=-\frac{\omega}{\omega_{0}}q-\frac{2\lambda Q}{\omega_{0}}\sqrt{1-\frac{1}{4}(Q^{2}+P^{2})}-\sqrt{\frac{2}{\omega_{0}}}\alpha,
\end{equation}

\begin{equation}
 \frac{\textrm{d}Q}{\textrm{d}t}=\frac{\partial H}{\partial P}=P-\frac{\lambda qQP}{2\omega_{0}\sqrt{1-\frac{1}{4}(Q^{2}+P^{2})}},
\end{equation}

\begin{equation}
 \frac{\textrm{d}P}{\textrm{d}t}=-\frac{\partial H}{\partial Q}=-Q+\frac{\lambda q Q^{2}}{2\omega_{0}\sqrt{1-\frac{1}{4}(Q^{2}+P^{2})}}-\frac{2\lambda q\sqrt{1-\frac{1}{4}(Q^{2}+P^{2})}}{\omega_{0}},
\end{equation}
subject to the initial conditions $q(t=0)=q_{0}$, $p(t=0)=p_{0}$, $Q(t=0)=Q_{0}$, and $P(t=0)=P_{0}$. Solving these equations yields the trajectory at any time, $\mathbf{x}(t)$, and its associated energy $H(\mathbf{x}(t))$ is always conserved and equal to the initial energy, $H(\mathbf{x}(t))=H(\mathbf{x}(0))$ for all $t$. Then one considers the intersection of this function with a given hyperplane at each time $t$; in this case we will intersect with $P=0$. When $\mathbf{x}(t)$ is such that $P(t)=0$, we collect the rest of phase coordinates and we finally represent the result in the $(p,q)$ plane.

\begin{figure*}[t]
\includegraphics[width=0.66\textwidth]{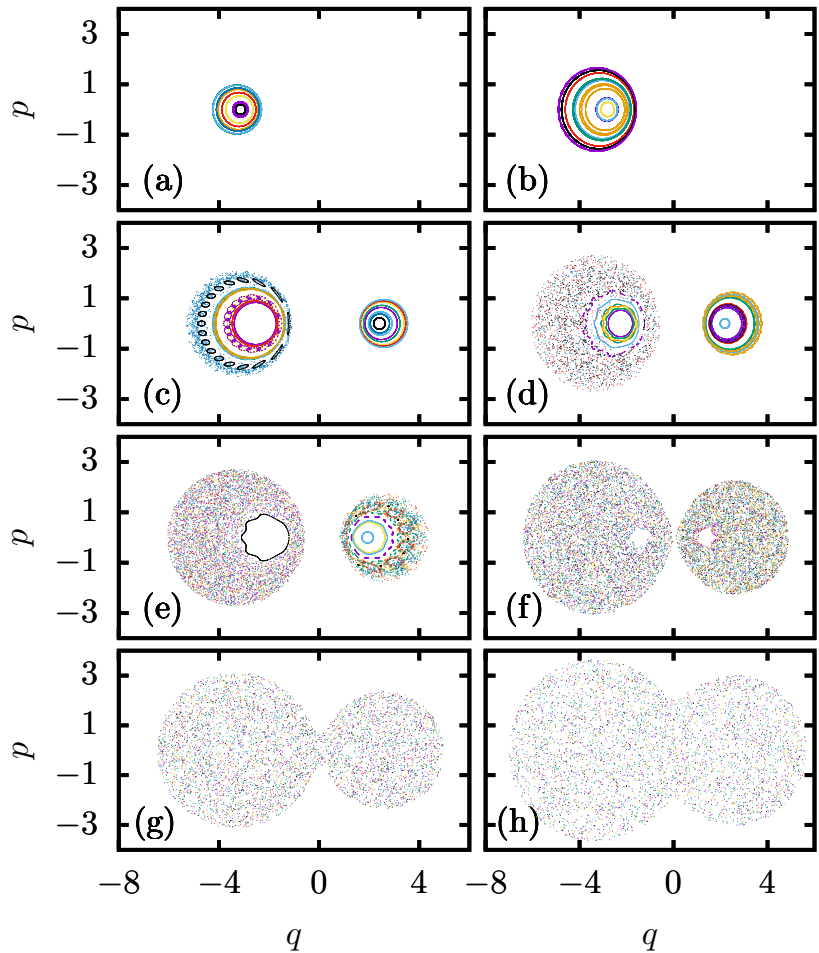}
\caption{Poincar\'{e} sections of the classical analogue Eq. \eqref{eq:semiclassical} in the $(p,q)$ plane. The trajectories have been interesected with the hyperplane $P=0$. Each panel corresponds to different energies of the trajectories: (a) $\epsilon=-5$, (b) $\epsilon=-4$, (c) $\epsilon=-3$, (d) $\epsilon=-2.5$, (e) $\epsilon=-2$, (f) $\epsilon=-1$, (g) $\epsilon=-0.8$, and (h) $\epsilon=1$. Each color corresponds to points of the same initial condition. }
\label{panelpoincare}
\end{figure*}

The two limiting behaviors of integrable and chaotic dynamics have clear characteristics. If the trajectory is regular, then its intersection with the hyperplane will produce ordered one-dimensional structures, often with the form of ovals (toroidal structures). By contrast, in the chaotic regime the trajectories can cover all the available phase space in an erratic way. 
In the transition between regularity and chaos regular portions of the phase space get distorted until they evolve into a fully disordered set of points as the regular tori are destroyed according to the KAM theorem \cite{Arnold1989}. 

The Poincar\'{e} sections thus obtained are represented in Fig. \ref{panelpoincare}. The points in each panel correspond to trajectories with a given fixed energy. At low energies, panels (a)-(b), the system is very approximately integrable, and we obtain an ordered set of points where each trajectory forms a toroidal curve. At these low energies there is a single structure in the Poincar\'{e} section since the second minimum does not exist yet. In panel (c) the energy is slightly above $\epsilon_{c1}$ where the second minimum appears. Interestingly, we observe that the left structure already shows some signatures of chaotic behavior in its outer region. However, the second structure on the right is completely regular, indicating that this part of the phase space is in fact approximately integrable and shows no trace of chaos. This is best exemplified in panel (d), where the left energy well is almost fully chaotic but the right well is still regular. In the rest of panels (e)-(h), we observe that as the energy of the picked initial condition is increased, chaos starts developing also in the right, previously regular well. When the energy is sufficiently high [panels (f)-(h)], chaos has completely won over regularity, and one has an ergodic mesh of points in the Poicar\'{e} sections. 

These results indicate that, at least in the classical limit of the quantum model, the addition of the perturbation parameter $\alpha$ has an important impact on the development of chaos. Namely, unlike in the standard Dicke model ($\alpha=0$) \cite{Bastarrachea2014b}, chaos  not only depends on the energy of the trajectories, 
 but also on whether the trajectory is attached to the left or the right energy well.
Thus, chaos develops in each well in an independent way. 

\section{Chaos in the quantum model}\label{sec:quantum}

We now turn to the quantum version of the model, Eq. \eqref{eq:hamiltonian}. The results of the previous section show that the onset of classical chaos depends on additional information than just the energy of the classical trajectory. The effect of the perturbation $\alpha$ is to break the symmetry of the classical energy wells, and chaos is not blind to this asymmetry. One then wonders whether similar features can also be observed in the quantum realm, i.e., whether chaos depends on additional conserved quantities besides energy. 

\subsection{Peres lattices}\label{subsec:pereslattices}

To answer this question, we will use the so-called Peres lattices, discussed by Peres already in the 1980s \cite{Peres1984}.  The main idea behind Peres' approach is the following \cite{Stransky2009}. One chooses a generic, few-body observable $\hat{\mathcal{O}}$ and then considers its eigenstate (diagonal) expectation values with respect to a given Hamiltonian, $O_{nn}\equiv \bra{E_{n}}\hat{\mathcal{O}}\ket{E_{n}}$. When represented as a function of the eigenenergies, these values $O_{nn}$ give rise to two kinds of limiting behaviors. An ordered set of points in the $(E_{n},O_{nn})$ plane (a `lattice') is an indication of regular dynamics, whereas a disordered set of points is associated with chaotic dynamics. In between these two limits, small perturbations of integrability can produce distortions of the regular lattice. Full chaos is only present when the entire structure of the lattice has been destroyed. Therefore, Peres lattices are an alternative way to study quantum chaos and integrability besides the more traditional analysis of level statistics. Additionally, Peres lattices in systems with two degrees of freedom have the advantage that, in a way, they can be thought of as a quantum counterpart of the classical Poincar\'{e} sections, and thus they are particularly well-suited for our work. Nonetheless, it should be noted that Peres lattices are not an exact quantum analogue of the classical Poincar\'{e} sections; however, much like Poincar\'{e} sections, Peres lattices do also provide a useful qualitative description of the emergence of chaos in the quantum model. 

\begin{figure*}[t]
\includegraphics[width=0.6\textwidth]{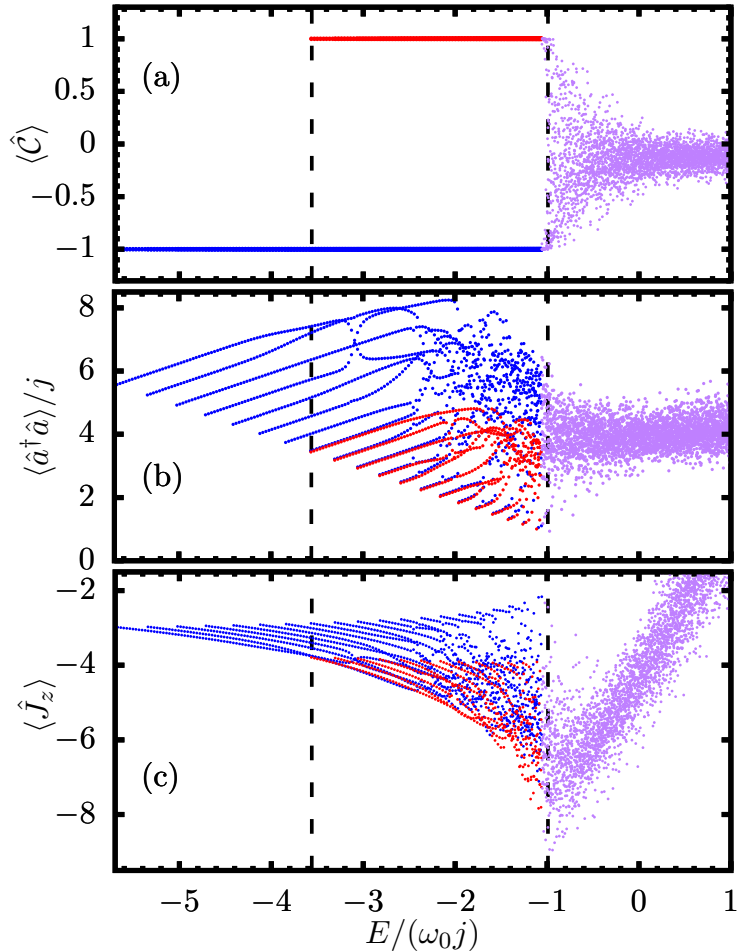}
\caption{Peres lattices of relevant observables in the eigenstates of the quantum Hamiltonian Eq. \eqref{eq:hamiltonian}. (a) Constant of motion Eq. \eqref{eq:constant}; (b) number of photons operator $\hat{a}^{\dagger}\hat{a}$; (c) atomic collective spin observable $\hat{J}_{z}$. Blue points show the expectation value in an eigenstate $\ket{E_{n}}$ for which $\bra{E_{n}}\hat{\mathcal{C}}\ket{E_{n}}=-1$, belonging to the left well of the classical phase space; red points correspond to $\bra{E_{n}}\hat{\mathcal{C}}\ket{E_{n}}=+1$, belonging to the right well; points for which $\hat{\mathcal{C}}$ is not a constant of motion are plotted in purple. Vertical dashed lines show the energy at which the right energy well is accessible, $\epsilon_{c1}=-3.563$, and the energy above which both energy wells are connected, $\epsilon_{c2}=-0.992.$}
\label{fig:panellattices}
\end{figure*}

To address the main question of this section, in conjunction with the Peres lattices we will need a mechanism to label the quantum eigenstates in relation to the properties of the asymmetric classical energy wells. To do so, we employ a constant of motion recently presented in \cite{Corps2021}. One can observe from the classical energy surfaces in Fig. \ref{fig:contoursDM} that below the second ESQPT critical energy, $\epsilon_{c2}$, the well structure (symmetric or asymmetric alike) is such that a classically conserved quantity can be defined. In particular, one can see that between the first and second ESQPTs,  $\epsilon_{c1}\leq\epsilon\leq \epsilon_{c2}$, two separate energy wells exist and they disconnected, meaning that a classical trajectory belonging to either well cannot pass over to the other well: each classical trajectory is trapped within only one of these two wells. For an energy value below the first ESQPT, $\epsilon_{GS}\leq \epsilon\leq \epsilon_{c1}$, there exists a single energy well, and therefore each trajectory is trivially attached to that well alone. 
However, above the second ESQPT critical energy, $\epsilon\geq \epsilon_{c2}$, the topology of the classical phase space does not allow for the above classification of trajectories since any initial condition can explore both regions of the phase space. The classical constant of motion below the second ESQPT is therefore actually a sign, $\textrm{sign}\,(q-q_{c})$, where $q_{c}=q_{2}^{*}$ is the classical coordinate corresponding to the second ESQPT critical energy $\epsilon_{c2}$ (i.e., the $q$ for which the curves appear to cross). It is possible to show \cite{Corps2021} that also in the quantum version of the model and in the thermodynamic limit, the quantum operator 

\begin{equation}\label{eq:constant}
\hat{\mathcal{C}}\equiv \textrm{sign}\,(\hat{q}-q_{c})
\end{equation}
is the corresponding quantum constant of motion. This constitutes a discrete $\mathbb{Z}_{2}$ symmetry, whose only two eigenvalues are $\textrm{Spec}\,\hat{\mathcal{C}}=\{-1,+1\}$. Thus, below the second ESQPT criticality, the diagonal expectation values of $\hat{\mathcal{C}}$ in the eigenstates of the Hamiltonian can only be $C_{nn}\equiv\bra{E_{n}}\hat{\mathcal{C}}\ket{E_{n}}\in\{-1,+1\}$ in the thermodynamic limit. Corrections to this behavior for finite sizes come from quantum tunneling between wells that is exponentially suppressed with system size \cite{Corps2021}. Still, for sufficiently high finite-$N$ values this is verified up to standard numerical precision [c.f. Fig. \ref{fig:panellattices}(a)]. These expectation values have a clear interpretation in terms of the geometry of the classical phase space. The value $-1$ is associated to the left energy well (where the ground-state belongs), while the $+1$ eigenvalue corresponds to the right energy well. Thus, $\hat{\mathcal{C}}$ assings a \textit{conserved quantum number} to each of the Hamiltonian eigenstates by which dynamics must abide. It tells us to which of the classical energy wells a given quantum eigenstate belongs. In the present work we are interested in revealing how this conservation law also has an impact on chaos.

Fig. \ref{fig:panellattices} shows the Peres lattices for several important observables. First, in panel (a) we represent the diagonal elements of $\hat{\mathcal{C}}$ defined above. We can observe that below the second ESQPT critical energy, $\epsilon_{c2}$, connecting both classical energy wells (marked by the second vertical dashed line), two values of $C_{nn}$ are possible, $+1$ and $-1$, in accordance with the arguments given above. Below the energy beyond which the right energy well is accessible (marked by the first vertical dashed line), $C_{nn}$ can only be $-1$, as there is only a single energy well. Colors in this figure are used to indicate the conserved charge provided by $\hat{\mathcal{C}}$: blue points correspond to eigenstates with $\langle\hat{\mathcal{C}}\rangle=-1$, while red points correspond to eigenstates with $\langle\hat{\mathcal{C}}\rangle=+1$. We can also see that above a certain energy threshold the expectation values show a disordered set of points, in purple, which is an indication of chaos. The energy above which chaos kicks in \textit{seems} to be related with the energy that connects both classical energy wells, but it is in fact unrelated: chaos appears below $\epsilon_{c2}$ even in the classical Poincar\'{e} sections [c.f. Fig. \ref{panelpoincare}(f)]. 

It is also worthwhile to point out to the effects of this panel (a) on thermalization. Quantum thermalization \cite{Peres1984b,Tasaki1998,Reimann2008,Deutsch2018,Alessio2016}, understood as in thermalization of individual eigenstates, refers to the phenomenon by which the long-time average of physical observables reaches an effective equilibrium state. This long-time average then simply fluctuates around this equilibrium value, which coincides with a suitable microcanonical average over a relevant energy width. When thermalization takes place, the infinite-time average approaches this microcanonical average as the system size increases, leading to the thermodynamic limit. The cornerstone of our understanding of quantum thermalization is summarized in the so-called eigenstate thermalization hypothesis (ETH) \cite{Alessio2016,Deutsch2018}, and it acknowledges quantum chaos as the main mechanism for why thermalization is possible. Simply put, if the system under consideration is quantum chaotic, its eigenstates will behave as
the eigenstates of random matrices,
and thus the diagonal expectation values of an observable will be a smooth function of energy. This is clearly incompatible with the expectation value $\langle\hat{\mathcal{C}}\rangle$, which shows discrete, abrupt jumps from $-1$ to $+1$ and viceversa as a function of energy. Thus, to study thermalization in these broken-parity Dicke models, one would need to consider two subsets of the spectrum, each with a definite value of $\langle\hat{\mathcal{C}}\rangle$: the dynamics of non-equilibrium states must obey the conservation law established by $\hat{\mathcal{C}}$ \cite{Corps2021}. An application of the generalized Gibbs ensemble may be better suited to treat this problem \cite{Vidmar2016}. This is in stark contrast with the standard version of the Dicke model. 

With this in mind, we move on to panels (b) and (c), which represent the Peres lattice of other two physically relevant observables. Panel (b) shows the case for the number of photons, $\langle\hat{a}^{\dagger}\hat{a}\rangle$. We can observe that at low enough energies, the diagonal expectation values form a completely regular lattice of blue points, indicating quantum regularity. For example, at $\epsilon=-5$, just slightly above the ground-state energy, this picture shows that there is apparently a strong correspondence between the quantum case and the Poicar\'{e} section represented in panel (a) of Fig. \ref{panelpoincare}, whose trajectories are also regular. The same applies for energy $\epsilon=-4$, which corresponds to panel (b) of Fig. \ref{panelpoincare}. We now focus on energy $\epsilon=-3$. We can see that at this energy the right energy well is already accessible. The red points corresponding to this energy well show again a regular pattern, close to the `ground-state' of the right well. However, the blue points, associated to the left well which already existed before, show some deviations from full regularity. This means that integrability is a little perturbed at this energy but this perturbation is not enough to completely destroy the lattice pattern. Interestingly, this can also be observed in panel (c) of Fig. \ref{panelpoincare}, where we observe that chaos has already started to appear but that there are strong remnants of regularity. This picture becomes more pronounced at energy $\epsilon=-2.5$, where we observe that the red quantum expectation values are essentially still regular, but the blue ones have now an important perturbation, with most points being disordered now. This is in concert with panel (d) of Fig. \ref{panelpoincare}, whose left well now shows a very small regular portion completely surrounded by a chaotic sea; meanwhile, the right well continues to show strong signatures of regularity. At energy $\epsilon=-2$, the blue expectation values are almost fully chaotic, and the red ones, although less chaotic than the blue, already show a strong perturbation of regularity. Panel (e) of Fig. \ref{panelpoincare} provides classical corroboration of this quantum picture. As energy is increased beyond $\epsilon=-2$, chaos starts to win over in both wells, until full chaos eventually kicks in. The same qualitative scenario is observed for the atomic observable $\hat{J}_{z}$, indicating that this result is independent of the part of the Hilbert space (atomic or photonic) one looks at.

These results show that in this deformed version of the Dicke model the addition of a term perturbating the symmetry of the energy wells is transferred onto the quantum properties of the model, in particular it significantly affects quantum chaos. Besides energy, quantum chaos also depends on an additional conserved quantity, given by the value of $\hat{\mathcal{C}}$ in each Hamiltonian eigenstate.

\subsection{Integrability breaking at low energies}\label{subsec:integrabilitybreaking}
A visible feature in the Peres lattices of Fig. \ref{fig:panellattices}(b)-(c) is the appearance of a set of independent points at low energies, from the ground-state up to some intermediate region of the spectrum. This band structure will allow us to provide a quantitative description of the onset of quantum chaos. 

\begin{center}
\begin{figure*}[t!]
\includegraphics[width=0.53\textwidth]{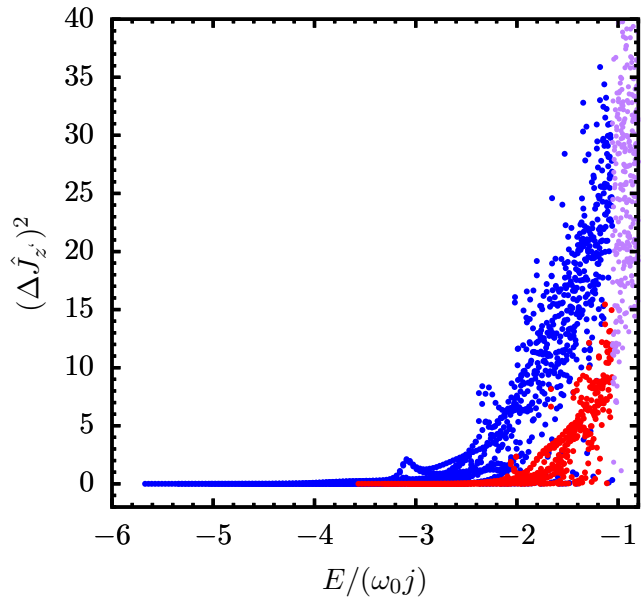} 
\caption{Dispersion Eq. \eqref{dispersion} of the adiabatic invariant $\hat{J}_{z'}$ Eq. \eqref{adiabaticinvariant} as a function of energy. Color code is the same as in Fig. \ref{fig:panellattices}. For the eigenstates belonging to both classical energy wells $(\Delta \hat{J}_{z'})^{2}=0$ at low energies with respect to the minimum energy of each well, meaning that at those energies each well behaves as an set of independent integrable models. Chaos sets in as $(\Delta \hat{J}_{z'})^{2}\neq0$. The energy value at which this occurs is different in each of the wells. The number of atoms is $N=60$.  }
\label{fig:deltaoppboa}
\end{figure*}
\end{center}

Even though the Dicke model is non-integrable, it has been shown \cite{Relano2016,Bastarrachea2017} that in the low-energy region it can be considered as very approximately integrable. This integrable region is characterized by an additional, adiabatic constant of motion $\hat{J}_{z'}$, which allows to divide the spectrum of the Dicke model into a set of $2j+1$ independent bands labelled by its eigenvalues. Using a Born-Oppenheimer approximation \cite{Berry1993}, one considers that the motion of the atoms is much faster than that of the bosons, so one freezes the slow bosonic coordinates and solves for them. This is accomplished by replacing the boson creation and annihilation operators by the simple semiclassical approximation $\hat{a}=(1/\sqrt{2})(\hat{q}+i\hat{p})$. Using the ideas of Ref. \cite{Relano2016}, the effective semiclassical Hamiltonians can be written

\begin{equation}\label{bands}
E_{m'}=\frac{\omega}{2}(p^{2}+q^{2})+\omega_{0}\sqrt{1+\frac{\lambda^{2}}{\lambda_{c}^{2}}\frac{\omega q^{2}}{\omega_{0}j}}m'+\frac{2}{\sqrt{\omega_{0}}}\alpha q,
\end{equation}
where the $2j+1$ quantum numbers $m'=-j,-j+1,\ldots,j-1,j$ are the eigenvalues of the adiabatic invariant 

\begin{equation}\label{adiabaticinvariant}
\hat{J}_{z'}\equiv \frac{\hat{J}_{z}+\sqrt{\frac{\lambda^{2}}{\lambda_{c}^{2}}\frac{\omega}{\omega_{0}}\frac{1}{2j}}(\hat{a}+\hat{a}^{\dagger})\hat{J}_{x}}{\sqrt{1+\frac{\lambda^{2}}{\lambda_{c}^{2}}\frac{\omega}{\omega_{0}}\frac{1}{2j}(\hat{a}+\hat{a}^{\dagger})^{2}}},
\end{equation}
which behaves as a constant of motion up to not too high energies \cite{Relano2016}. Eq. \eqref{bands} indicates that at low energies the Dicke model, which has two semiclassical degrees of freedom as shown in Eq. \eqref{eq:semiclassical}, can be split into a set of independent integrable models with a single effective degree of freedom each. Each $E_{m'}$ defines a different classical energy surface; the ground-state belongs to the surface with $m=-j$. It follows that in the quantum case the energy levels depend on two quantum numbers, $E_{m',n}$: $m'$ indicates to which band a given energy level belongs, while $n$ indexes its position relative to the minimum energy within that band. The diagonal expectation values $\bra{E_{n}}\hat{J}_{z'}\ket{E_{n}}\in\{-j,-j+1,\ldots,j-1,j\}$ can be used to classify the quantum eigenlevels in bands. 

It should be noted that the adiabatic invariant $\hat{J}_{z'}$ Eq. \eqref{adiabaticinvariant} and the constant of motion $\hat{\mathcal{C}}$ Eq. \eqref{eq:constant} are independent and fundamentally unrelated. Their physical meaning is different: $\hat{J}_{z'}$ indicates to which low-energy band a given eigenstate belongs, while $\hat{\mathcal{C}}$ indicates within which classical energy well a given eigenstate is trapped. These are not mutually exclusive, as two states corresponding to different values of $\hat{\mathcal{C}}$ can nonetheless be found in the same energy band as indexed by $\hat{J}_{z'}$. Moreover, the regions of the spectrum where \textit{both} operators, $\hat{J}_{z'}$ and $\hat{\mathcal{C}}$, act as good constants of motion have a non-empty intersection: $\hat{\mathcal{C}}$ is a constant of motion up to the ESQPT critical energy connecting the classical energy wells \cite{Corps2021}, while $\hat{J}_{z'}$ is valid up to some energy value that depends on the coupling parameters of the model \cite{Relano2016}. At intermediate energies below the second ESQPT and at low energies near the ground-state, both operators act simultaneously as constants of motion, providing two independent quantum numbers, $\langle \hat{J}_{z'}\rangle$ and $\langle\hat{\mathcal{C}}\rangle$, labelling the system states. It is interesting to note that $[\hat{J}_{z'},\hat{\mathcal{C}}]=0$ since both operators are diagonal in the basis diagonalizing $\hat{a}+\hat{a}^{\dagger}$.

It has been argued \cite{Relano2016,Bastarrachea2017} that the quantum chaotic domain emerges when the operator Eq. \eqref{adiabaticinvariant} ceases to be a constant of motion. We can use this fact to provide a more quantitative picture than that given by the Peres lattices, which are qualitative. As an indicator of quantum chaos, we consider the dispersion of $\hat{J}_{z'}$ in the eigenstates $\{\ket{E_{n}}\}_{n}$ of the quantum Hamiltonian, 

\begin{equation}\label{dispersion}
(\Delta \hat{J}_{z'})^{2}=\bra{E_{n}}\hat{J}_{z'}^{2}\ket{E_{n}}-(\bra{E_{n}}\hat{J}_{z'}\ket{E_{n}})^{2}\geq0.
\end{equation}
If this dispersion vanishes, then $\hat{J}_{z'}$ is a good conserved quantity. As this dispersion increases from zero, the integrability found at low energies is perturbed and the onset of chaos begins.

The dispersion Eq. \eqref{dispersion} is shown in Fig. \ref{fig:deltaoppboa}. As in Fig. \ref{fig:panellattices}, we have classified each quantum eigenstate according to the quantum number assigned by $\hat{\mathcal{C}}$, and thus blue and red points represent eigenstates belonging to the left, $\langle\hat{\mathcal{C}}\rangle=-1$, and right, $\langle\hat{\mathcal{C}}\rangle=+1$, classical energy wells, respectively. We can see that in both cases the dispersion $(\Delta \hat{J}_{z'})^{2}$ vanishes near the minimum energy of each of the two classical wells (each of the two `ground-states'), indicating that those low-lying eigenstates can be successfully classified in independent energy bands by $\hat{J}_{z'}$. Thus, near these two minima both wells behave as independent integrable systems. We now focus on the states belonging to the left energy well, with $\langle\hat{\mathcal{C}}\rangle=-1$. The vanishing dispersion is preserved up to $\epsilon\approx -3.5$, where $(\Delta \hat{J}_{z'})^{2}$ stops being zero and grows. This indicates that the classification of the eigenstates in energy bands is no longer feasible, this perturbation being due to the appearance of chaos. As the energy further increases, the dispersion continues to grow. A similar scenario is obtained for the states with $\langle\hat{\mathcal{C}}\rangle=+1$, whose dispersion near the energy minimum of that classical well is zero and then also starts to grow at $\epsilon\approx -2.5$. The energy value where the dispersion acquires a non-zero value is significantly higher than the analogous point for the states with $\langle\hat{\mathcal{C}}\rangle=-1$.

These results provide clear, quantitative evidence that chaos appears at quite different energies and that the dynamics of a classical energy well decouples from the dyamics of the other well. Below the second ESQPT critical energy, $\epsilon_{c2}$, the system effectively behaves as two independent models whose properties need to be analyzed separately. Above this energy, the distinction between wells no longer applies as both wells merge into a single one.

\section{Conclusions}\label{sec:conclusions}
In this work we have explored the development of both classical and quantum chaos in a version of the Dicke model of quantum optics. By introducing a deformation strength into the standard Hamiltonian, in the semiclassical limit one obtains two non-equivalent, asymmetric energy wells at different energies. Below a certain energy threshold, intimately associated with an excited-state quantum phase transition, classical trajectories are trapped within each of the wells and cannot pass over to the other one. Above the critical energy of the second ESQPT, the two wells become connected. By means of classical Poincar\'{e} sections we have shown that each of the two wells are characterized by independent dynamics: at the same energy, depending on which well a trajectory belongs it can be either regular or fully chaotic. In the quantum realm, Peres lattices of representative atomic and photonic observables reveal an interesting connection with the classical case. The operator $\hat{\mathcal{C}}$ acts a conserved quantity below the second ESQPT critical energy and assigns a conserved quantum number, $\pm 1$, to each of the Hamiltonian eigenstates depending on which of the two classical energy wells they are trapped. The development of chaos is then shown to depend strongly on the conservation law imposed by $\hat{\mathcal{C}}$. As a consequence, the usual theory of quantum thermalization is invalid to describe the long-time averaged equilibrium states reached in this family of systems. For suitable values of the control parameters and at certain energy values, one can thus have a system in which chaos and regularity coexist in a completely independent fashion. The emergence of the chaotic domain has been further analyzed by taking into account the low-energy band structure of the model. The energy at which this approximate integrability is broken coincides rather well with quantum Peres lattices and classical Poincar\'{e} sections. Our results could be experimentally tested in trapped ions in a linear trap \cite{Aedo2018}. An applied constant (in the reference frame rotating with the ion lattice) electric field can induce the parity breaking term in the Hamiltonian (\ref{eq:hamiltonian}). The classical dynamics could be also simulted with electrical circuits \cite{Quiroz2020}.

\acknowledgements
This work has been financially supported by the Spanish Grant No. PGC2018-094180-
B-I00 (MCIU/AEI/FEDER, EU), CAM/FEDER Project No. S2018/TCS-4342 (QUITEMAD-CM),
and CSIC Research Platform on Quantum Technologies PTI-001. A. L. C. acknowledges financial support from `la Caixa' Foundation (ID 100010434) through the fellowship LCF/BQ/DR21/11880024.

\end{document}